\documentclass[12pt]{article}
\usepackage{graphicx, epsfig}
\usepackage{amsmath}
\usepackage{latexsym}
\usepackage{amstext}
\usepackage{ulem}
\usepackage{amssymb}
\usepackage{array}
\usepackage{multirow}
\usepackage{tikz}
\newcommand{\ds}{\displaystyle}
\newcommand{\beq}{\begin{eqnarray}}
\newcommand{\eeq}{\end{eqnarray}}
\newcommand{\beqq}{\begin{eqnarray*}}
\newcommand{\eeqq}{\end{eqnarray*}}

\begin{document}

\pagestyle{plain}

\begin{center}

{\bf The case of  escape probability as linear in short time  }\\[5MM]

A. Marchewka\\[3mm]
8 Galei Tchelet St., Herzliya, Israel\\
e-mail avi.marchewka@gmail.com,\\[5mm]
Z. Schuss\\[5mm]
Department of Mathematics\\[0pt]
Tel-Aviv University\\[0pt]
Ramat-Aviv, Tel-Aviv 69978, Israel\\[0pt]
e-mail schuss@math.tau.ac.il

\vspace{1cm} \vspace{3mm} {\bf ABSTRACT}

\end{center}

\vspace{5mm}
We derive rigorously the short-time escape probability of a quantum particle from its compactly
supported initial state, which has a discontinuous derivative at the boundary of the support. We
show that this probability is linear in time, which seems to be a new result. The novelty of our
calculation is the inclusion of the boundary layer of the propagated wave function formed outside
the initial support. This result has applications to the decay law of the particle,
to the Zeno behaviour, quantum absorption, time of arrival, quantum measurements, and more.

\pagestyle{plain} \setcounter{page}{1}

\section{Introduction}
There are two types of propagation laws that a quantum particle obeys: unitary, as given by the
Schr\"{o}dinger evolution, and non-unitary, as asserted by the collapse axiom. In each of these
laws hides a puzzling ingredient.  The reconciliation of these apparently incompatible
laws poses a big challenge in every day physics. Indeed, a related problem to is the Zeno effect,
time of arrival, absorption phenomena, and so on. A formulation of an irreversible decay law of
the initial state can be a useful way to link between the two types of evolution \cite{MSA}.
Clearly, decay laws are often derived from the short-time behaviour of the escape probability from the
initial state \cite{MUG}. The purpose of this paper is to derive a new role of the escape
probability  in spatial coordinates. The difficulty in the formulation of such a law stems from
the short-time evolution of the wave function, that is, from the short-time escape probability
from the initial state.

To clarify the problem, we set first $\hbar=1,\ 2m=1$ and consider the time-evolution of a
spatial initial state $\left| \psi\right\rangle$, given by
\begin{align}
\left| {\psi \left( t \right)} \right\rangle  = {e^{ - i\hat Ht}}\left| \psi  \right\rangle,
\label{0}
\end{align}
where it suffices for our purpose to consider only the free-particle evolution $\hat H={\hat
p}^2$ with a time-independent Hamiltonian of the system, and the initially normalized state
$|\psi\rangle$. We assume  $|\langle x |\psi(t)\rangle|^2\neq|\langle x|\psi\rangle|^2$, which means that the initial state propagates. The usual short-time expansion is \cite{Peres}
\begin{align}\label{11}
\left| {\psi \left( t \right)} \right\rangle  = \left( {1 - i\hat Ht-t^2\frac{\hat{H}^2}{2}} \right)\left| \psi  \right\rangle  + O\left(t^3\right),
\end{align}	
which gives the escape probability
\begin{align}\label{12}
E\left( t \right) \equiv &\,1- {\left| {\left\langle \psi  \right|\left| {\psi \left( t \right)} \right\rangle } \right|^2}
 \simeq {\left| {\left\langle \psi  \right|\left| \psi  \right\rangle }
 \right|^2}it  {\left\langle \psi  \right|\left( {\hat H - {{\hat H}^*}} \right)
 \left| \psi  \right\rangle }  \nonumber\\
 &- {t^2}\left\langle \psi  \right|\hat H\left| \psi  \right\rangle \left\langle \psi  \right|{{\hat H}^*}\left| \psi  \right\rangle
 +t^2\langle\psi|\psi\rangle\langle\psi|\hat{H}^2|\psi\rangle+O(t^3).
\end{align}	  	
Eq.\eqref{12} for a self-adjoint  Hamiltonian operator (symmetric and bounded), $\hat H^{\dagger} =\hat H$,  implies that for $t\to0$
\begin{align}\label{teps}
E\left( t \right) =  {t^2}\Delta H+O(t^3),
\end{align}	
where  $\Delta H=\langle\psi|H|\psi\rangle^2-\langle\psi|H^2|\psi\rangle$.

This law means that the initial state does not decay under continuous observations.
To see this, we consider a sequence of projection measurements on a spatial
interval $\Omega$ at times $t_i=i\Delta t$, such that $T=N\Delta t$. If the probability
to measure  at each time $i\Delta t$ is given by \eqref{teps}, then the
probability to find the particle after $N$ measurements is approximately by the upper bound 
 \begin{align}
 E(N\Delta t)\approx {\left(\frac{T}{N}\right)}^2\sum\limits_{i=1}^{N}\Delta H_i\leq \frac{T^2}{N}\max\Delta H_i,\label{Qudrtic}
 \end{align}
where $\Delta H_i=\langle\psi(t_{i-1})|H|\psi(t_{i-1})\rangle^2-\langle\psi(t_{i-1})|H^2|\psi(t_{i-1})\rangle$.
 The escape probability, that is,
the probability to detect the particle by time $T$, is  converged to zero as $\Delta t\to0$.
This means that in this model of measurements, the decay of the particle is
inhibited, which is a peculiar result from the classical point of view; this inhibition of decay
is called the Zeno paradox (see \cite{Peres}, \cite{Home}).
Had the time \eqref{12} been linear rather then quadratic, the limit in \eqref{Qudrtic} would be finite and not zero and thus the decay would be uninhibited in this model. The demonstration of the linear short-time behaviour is the main result of this paper.

The cause of the failure of \eqref{teps} is the generally agreed model of continuous measurement, which under \eqref{teps} consists of frequently repeated
instantaneous detections. The condition \eqref{12} is necessary for the above formalism to be valid. However, it is valid universally for
all initial states only for bounded (continuous) linear operators $\hat A$, such that for all functions $\psi$ in their domains
\begin{align}\label{13}
\left\langle \psi  \right|\hat A\left| \psi  \right\rangle  \le C\left\langle \psi  \right|\left| \psi  \right\rangle
\end{align}	
for some constant $C$. This is not the case for the unbounded Schr\"{o}dinger operator $\hat H$,
and thus \eqref{12} does not hold in all cases. Indeed, it fails for compactly supported wave
functions, whose derivative on the boundary of the support is undefined. This is exactly the point where the above derivation fails for the case of
initially discontinuous or non-smooth initial wave functions (in spatial coordinates). See \cite{Friedman} for details.

Indeed, the short-time escape probability of an initially non-smooth wave function has been
discussed several times in the literature \cite{MSA},\cite{MUG},\cite{PLA},\cite{GM},\cite{MOH}. Specifically, it has been shown that in the above
mentioned cases the type of discontinuity of the initial wave function determines its  escape probability.
It was shown by the steepest descent method that an initial wave function with a jump discontinuity at the boundary of its support (at $x=0$, say), has escape probability  into $x>\delta>0$ of the order
\begin{align*}
  E(\delta,\Delta t)=O\left(\frac{\Delta t}{\delta^2}\right)\quad\mbox{as}\ \Delta t\to0.
\end{align*}

This linearity in short time has some significance \cite{MOH}, but also some limitations \cite{PLA2}. Indeed, in repeated observations the wave function in
the initial support becomes continuous, but its derivative gains a jump at the boundary
 \cite{MSA}, \cite{PLA}, \cite{SCH}.
For a continuous initial wave function $\psi(x,0)$ with a discontinuous derivative at the
boundary the steepest descent gives
\begin{align*}
  E(\delta,\Delta t)=O\left(\frac{\Delta t^3}{\delta^4}\right).
\end{align*}
The limitation of the large phase approximation of the propagated
wave function is its divergence in the boundary layer outside the support,
which is of the form $x/\sqrt\Delta t$.

Usually, the time-dependence of $E(\delta,\Delta t)$ is interpreted
as Zeno behaviour of the wave function \cite{MSA}, \cite{MUG}, \cite{SCH}. However, the
$1/\delta^2$ dependence is
problematic, because it implies the divergence of the escape probability.
Higher-order terms in the large phase approximation fail to recover convergence.
Thus the contribution of the boundary
layer has to be accounted for, which is the main result of this paper.

\section{Short-time escape probability}\label{ss:Decay}

We consider the escape probability of a particle with an initially continuous wave function $\psi_0(x)$, compactly supported in  $\Omega=(-1,0)$. Furthermore, $\psi_0(x)$ has a continuous derivative $\psi_0'(x)$ and jumps discontinuously across the boundary to the value $0$.

The free propagation of the initial wave function $\psi_0(x)$ in the time interval
$(0,\Delta t)$ is given explicitly in terms of Green's function for the Schr\"{o}dinger initial value problem as
\begin{align*}
\psi(x,\Delta t)=\frac{1}{\sqrt{\pi i \Delta t}}
\int\limits_{-1}^{0}\psi_0(y)\exp\left\{\frac{-i(x-y)^2}{\Delta t}\right\}dy.
\end{align*}
Approximating $\psi_0(y)\simeq \psi_0'(0)y $ and changing variable to $z=(x-y)/\sqrt{\Delta t},$
we get
\begin{align*}
\psi(x,\Delta t)=\frac{\psi_0'(0)}{\sqrt{i\pi}}\int\limits_{x/\sqrt{\Delta t}}^{(x+1)/\sqrt{\Delta t}}\left(x-\sqrt{\Delta t}
z\right)\exp\left\{\frac{-iz^2}{2}\right\}dz[1+o(1)]\hspace{0.5em}\mbox{as}\ \Delta t\to0.
\end{align*}
The integral of $\sqrt{\Delta t}z$ is the exact expression
\begin{align*}
&\frac{\psi_0'(0)\sqrt{\Delta t}}{\sqrt{i\pi}}\int\limits_{(x+1)/\sqrt{\Delta t}}^{x/\sqrt{\Delta t}}z\exp\left\{-\frac{iz^2}{2}\right\}dz\\
=&\frac{\psi_0'(0)\sqrt{\Delta t}}{-i\sqrt
{i\pi}}\left[\exp\left\{\frac{-ix^2}{2\Delta t}\right\}
-\exp\left\{\frac{-i(x+1)^2}{2\Delta t}\right\}\right].
\end{align*}
The escape probability into the positive ray in short time $\Delta t$ is given  in this case by  $$E(\Delta t)\equiv\int\limits\limits_{0}^{\infty}|\psi(x,\Delta t)|^2dx,$$
that is,
 \begin{align*}
E(\Delta t)=&\int\limits_{0}^{\infty}\left|\frac{-\psi_x(0,0)x}{\sqrt{i\pi}}
\int\limits_{(x+1)/\sqrt{\Delta t}}^{x/\sqrt{\Delta t}}\exp\left\{-\frac{iz^2}{2}\right\}dz \right.\\
&\left. +\frac{\sqrt{it}\psi_x(0,0)}{\pi}\left[\exp\left\{\frac{-ix^2}{2\Delta t}\right\}-\exp\left\{\frac{-i(x+1)^2}{2\Delta t}\right\}\right]\right|^2dx[1+o(1)].
 \end{align*}
Changing $\xi=x/\sqrt{\Delta t}$, we get
 \begin{align*}
  E(\Delta t)=&\frac{\Delta t^{3/2}|\psi_x(0,0)|^{2}}{\pi}\int\limits_{0}^{\infty}\left|\frac{\xi}{\sqrt{i}}
\int\limits_{\xi}^{\xi+1/\sqrt{\Delta t}}\exp\left\{-\frac{iz^2}{2}\right\}dz\right.\\ &\left.+\sqrt{i}\left[\exp\left\{\frac{-i\xi^2}{2}\right\}-\exp\left\{\frac{-i\{\xi\sqrt{\Delta t}+1)^2}{2\Delta t}\right\}\right]\right|^2\,d\xi[1+o(1)].
 \end{align*}
To estimate the inner integral, we change $\eta=z^2$, which gives
 \begin{align}\label{}
\int\limits_{\xi}^{\xi+1/\sqrt{\Delta t}}\exp\left\{-\frac{iz^2}{2}\right\}dz=\int\limits_{\xi^2}^{(\xi+1/\sqrt{\Delta t})^2}\frac{\exp\left\{-\ds\frac{i\eta}{2}\right\}}{2\sqrt{\eta}}d\eta.
 \end{align}
Now, we break   $$\int\limits_{\xi^2}^{(\xi+1/\sqrt{\Delta t})^2}=
\int\limits_{\xi^2}^{\infty}-\int\limits_{(\xi+1/\sqrt{t})^2}^{\infty}$$
and estimate each part separately. In short time $\Delta t$ (i.e., large $\xi$),  we find
that
 \begin{align*}
&\frac{\xi}{2\sqrt{i}}\left(\int\limits_{\xi^2}^{\infty}-
\int\limits_{(\xi+1/\sqrt{t})^2}^{\infty}\right)\frac{\exp\left\{-\ds\frac{i\eta}{2}\right\}}
{\sqrt{\eta}}d\eta\\
=&\sqrt{i}\xi\left[\frac{\exp\left\{-\ds\frac{i(\xi+1/\sqrt{t})^2}{2}\right\}}
{\xi+1/\sqrt{\Delta t}}-\frac{\exp\left\{-\ds\frac{i\xi^2}{2}\right\}}{\xi}\right][1+o(\Delta t)].
 \end{align*}
 Using this result for the escape probability, we find that
 \begin{align}
   E(\Delta t)=\frac{{\Delta t}^{3/2}|\psi_0'(0)|^{2}}{\pi}
   \int\limits_{0}^{\infty}\left|\frac{\exp\left\{-\ds\frac{i(1+\xi\sqrt{\Delta t})^2}{2\Delta t}\right\}}{1+\xi\sqrt{\Delta t}}\right|^2\,d\xi=\frac{\Delta t |\psi_0'(0)|^{2}}{\pi},\label{EL}
 \end{align}
 where we have used the integral $$\int\limits_{0}^{\infty}\frac{1}{(1+\xi\sqrt{\Delta t})^2}d\xi=\frac{1}{\sqrt{\Delta t}}.$$
 Note that the contribution of the left boundary to the escape probability on the right is $O(\Delta t^{3/2})$, (see, e.g., \cite{MUG}, \cite{GM},
\cite{SCH}),  which is  negligible.
Adding to \eqref{EL} the escape probability into the ray $x<-1$, we obtain the total escape probability
\begin{align}\label{main}
E(\Delta t)=\frac{\Delta t (|\psi_0'(0)|^{2}+|\psi_0'(-1)|^{2})}{\pi}[1+O(\sqrt{\Delta t})].
\end{align}
Note that the escape probabilities in the different directions are not the same. Eq.\eqref{main} is the main results of this paper.

\begin{figure}
  \centering
  \includegraphics[width=5cm]{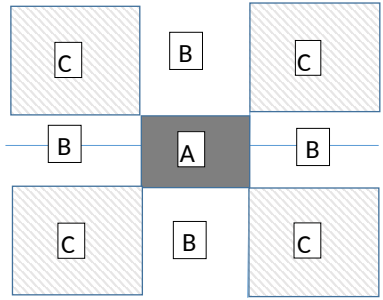}
\caption{\small The square A is the initial support. The infinite strips B are the domains of linear short-time escape probability. The quadrants C are the domains of escape probability $O(\Delta t^2)$ in short time}\label{Fig:figure1}
\end{figure}

Consider, for example,  the two-dimensional compactly supported wave function in a square. The propagator in the plane is given now by
 \begin{align*}
   \psi(x,y,\Delta t)=\frac{1}{\pi i \Delta t}\int\limits_{-1}^{0}\int\limits_{-1}^{0}
   \exp\left\{\frac{-i(x-\xi)^2}{\Delta t}+\frac{-i(y-\eta )^2}{\Delta t}\right\}\varphi(\xi,0)\phi(\eta,0)\,d\xi\,d\eta
 \end{align*}
 For an internal point $x\in (-1,0)$ and $y\notin(-1,0)$, we have
 \begin{align*}
   \psi(x,y,\Delta t)=&\frac{\varphi(x,0)}{\sqrt{\pi i \Delta t}}\int\limits_{-1}^{0}
   \exp\left\{\frac{-i(y-\eta )^2}{\Delta t}\right\}\phi(\eta,0)\, d\eta[1+O(\Delta t)]\\
 =&O(\sqrt{\Delta t}).
 \end{align*}
 Therefore, we find that the escape probability to the domain $B$ (see in Fig.\ref{Fig:figure1}) is linear in short time, but  it is $O(\Delta t^{2})$ in the domain $C$, in which both $x$ and $y$ are outside the interval. This is a known wave phenomenon in classical waves that propagate along characteristics.

\section{Summary and discussion}
The result of this paper is the derivation of the short-time linear escape probability
from first principles, which is a new result for short times. The
significance of this result is in its application to the Zeno behaviour,
continuous observations, Zeno dynamics, and more. For example,
it is well understood that exponential of decay law is linear in time.
However, it is generally agreed that in short time the decay of quantum states
is quadratic, e.g. \cite{PFacchi} and hence the possibility
of exact exponential decay is excluded. This also can be seen from equation \eqref{Qudrtic}.
Therefore, the result \eqref{main} is an exception to the widely agreed quadratic
short-time behaviour of systems, which  leads to an exact exponential decay law.
The linear short time behavior \eqref{main} is
implied from the initially continuous wave function $\psi_0(x)$,
compactly supported in $\Omega$ with a continuous derivative $\psi_0'(x)$ and
jumps discontinuously across the boundary.
Possible realizations of such a quantum state are:
an initial neutron bound inside a nucleus, where its decay product, (say $\beta ^-$ ),
is the escape probability, and also the releasing from a trap scenario where the trap's state is the initial
such a bounded wave function. The escape from the trap is therefore the escape probability.

To our knowledge, there is no other example that exhibits a linear escape probability in short
time. The full construction of the physical setup of exponential-type decay will be done elsewhere.

\end{document}